\documentclass[]{emulateapj}
\usepackage{graphicx}
\usepackage{latexsym}
\usepackage{natbib}

\begin{document}

\title{Intermittent millisecond X-ray pulsations from the neutron-star X-ray transient  SAX~J1748.9--2021 in the globular cluster NGC~6440}

\author{D. Altamirano\altaffilmark{1}, 
P. Casella\altaffilmark{1}, 
A. Patruno\altaffilmark{1},   
R. Wijnands\altaffilmark{1},  
M. van der Klis\altaffilmark{1}}

\altaffiltext{1}{Email: diego@science.uva.nl ; Astronomical Institute, ``Anton Pannekoek'',
University of Amsterdam, and Center for High Energy Astrophysics,
Kruislaan 403, 1098 SJ Amsterdam, The Netherlands.}

\begin{abstract}

We report on intermittent X-ray pulsations with a frequency of
442.36~Hz from the neutron-star X-ray binary SAX~J1748.9--2021 in the
globular cluster NGC~6440. The pulsations were seen during both 
2001 and 2005 outbursts of the source, but only intermittently,
appearing and disappearing on timescales of hundreds of seconds.
We find a suggestive relation between the occurrence
of type-I X-ray bursts and the appearance of the pulsations but the relation
is not strict.
This behavior is very similar to that of the intermittent accreting
millisecond X-ray pulsar HETE~J1900.1--2455. The reason for the
intermittence of the pulsations remains unclear.
However it is now evident that a strict division between pulsating and
non-pulsating does not exist.
By studying the Doppler shift of the pulsation frequency we determine
an orbit with a period of 8.7 hrs and an projected semi major axis of
0.39 lightsec. The companion star might be a main--sequence or a
slightly evolved star with a mass of $\sim$1 $M_\odot$.
Therefore, SAX~J1748.9--2021 has a longer period and may have a more
massive companion star than all the other accreting millisecond X-ray
pulsars except for Aql X-1.

\end{abstract}
\maketitle

\section{Introduction}
\label{sec:intro}

Accreting millisecond pulsars \citep[AMPs,][]{Alpar82,Backer82} are
transient low mass X-ray binaries (LMXBs) that show X-ray pulsations
during their outbursts. A total of nine AMPs out of ~100 non-pulsating
LMXBs have been found to date. The reason why only this small subgroup
of binaries pulsates is still unknown. The first seven AMPs discovered
showed persistent X-ray pulsations throughout the outbursts. Recently
\citet{Kaaret06} discovered the AMP HETE~J1900.1--2455, which has
remained active for more than 2 years\footnote{At the time of
submitting this letter, the source is still active.} but showed
pulsations only intermittently during the first $\sim2$ months of
activity \citep{Galloway07a}.
From the transient source Aql X-1 pulsations were detected
\citep{Casella07} only for $\sim150$~sec out of the $\sim1.3$~Msec the
source has (so far) been observed with the Rossi X-ray Time Explorer
(RXTE).

\citet{Gavriil06,Gavriil07} recently reported on the detection of
$\sim442.3$~Hz pulsations in an observation of the 2005 outburst of a
transient source in the globular cluster (GC) NGC~6440. The pulsations
followed a flux decay observed at the beginning of the observation and
were reminiscent of those observed during superbursts; however, as
\citet{Gavriil07} suggest, they could also be a detection from a new
intermittent accreting millisecond pulsar.  \citet{Kaaret03} report a
409.7~Hz burst oscillation in an X-ray transient (SAX~J1748.9--2021)
located also in NGC~6440 and this GC harbors at least 24 X-ray sources
\citep{Pooley02}, so \citet{Gavriil07} concluded that the burst
oscillations and the pulsations were probably coming from different
X-ray transients in the same GC.

The exact formation mechanisms behind the pulsations of these three
sources remains unknown. 
The existence of intermittent pulsations with a small duty cycle
implies that many other apparently non-pulsating LMXBs might be 
pulsating, bridging the gap between the small number of AMPs
and the large group of non-pulsating LMXBs. 

We are performing a detailed analysis of all RXTE archival data of
neutron-star LMXBs to search for transient pulsations in their X-ray
flux \citep[see also][]{Casella07}. 
In this Letter we present the results of our search on the three X-ray
outbursts observed from the globular cluster NGC~6440.

\section{The neutron-star transient SAX~J1748.9--2021 in NGC~6440}

NGC~6440 is a GC at $8.5\pm0.4$~kpc \citep{Ortolani94}. Bright X-ray
outbursts from a LMXB were reported in 1971, 1998, 2001 and 2005
\citep[]{Markert75,Zand99,Verbunt00,Zand01,Markwardt05}. 
\citet{Zand01} from X-ray and optical observations concluded that the
1998 and 2001 outbursts were from the same object, which they
designated SAX~J1748.9--2021.

\begin{figure*}[t] 
\center
\resizebox{2\columnwidth}{!}{\rotatebox{0}{\includegraphics{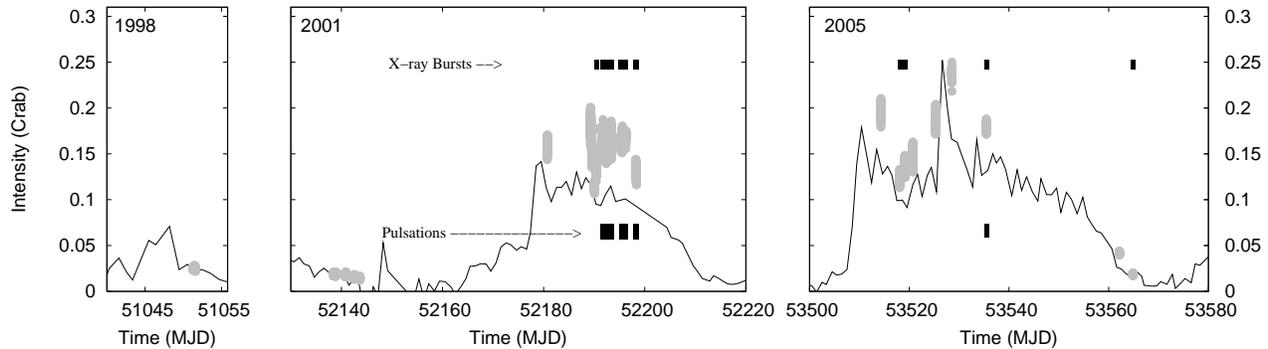}}}
\caption{Intensity (2.0--16.0~keV) normalized by Crab vs. time of the
three outbursts. Gray symbols show the 16-sec averaged intensity
during the pointed PCA observations.
The continuous line shows the ASM light curve.
Black marks at the \textit{top} mark the times of type-I X-ray
bursts. Black marks at the \textit{bottom} mark the times when we
detect significant pulsations. Years of the outburst is shown in each
panel. }
\label{fig:lc}
\end{figure*}

\section{Observations, data analysis and results}
\label{sec:dataanalysis}

We used data from the RXTE Proportional Counter Array \citep[PCA, for
instrument information see][]{Jahoda06}. Up to July, 2007, there were
27 pointed observations of SAX J1748.9--2021, each covering 1 to 5
consecutive 90-min satellite orbits. Usually, an orbit contains
between 1 and 5 ksec of useful data separated by 1--4 ksec data gaps
due to Earth occultations and South Atlantic Anomaly passages.
Adopting a source position \citep[$\alpha=17^h 48^m 52^s.163$, $\delta
= -20^o 21^{'} 32^{''}.40$; J2000][]{Pooley02} we converted the
2--60~keV photon arrival times to the Solar System barycenter with the
FTOOL faxbary, which uses the JPL DE-200 ephemeris along with the
spacecraft ephemeris and fine clock corrections to provide an absolute
timing accuracy of ~5-8 $\rm\,\mu s$ \citep{Rots04}.

We performed a Fourier timing analysis using the high-time resolution
data collected in the Event (E\_125us\_64M\_0\_1s) and the Good Xenon
modes.  Power spectra were constructed using data segments of 128, 256
and 512 seconds and with a Nyquist frequency of 4096~Hz.  No
background or dead-time corrections were made prior to the calculation
of the power spectra, but all reported rms amplitudes are background
corrected; deadtime corrections are negligible.

\begin{table}
\caption{Timing parameters for NGC~6440}

\scriptsize
\begin{tabular}{lc}
\hline
\hline
Parameter & Value \\
\hline
Orbital period, P$_{orb}$(hr) \dotfill                           & 8.764(6) hr \\
Projected semi major axis, $a_x sin i$ (lightsec.)\dotfill   & 0.39(1) \\
Epoch of 0$^o$ mean longitude$^1$, $T_0$ (MJD/TDB) \dotfill          & 52190.047(4) \\
Eccentricity, e \dotfill                                         & $<0.001$ \\
Spin frequency $\nu_0$ (Hz) \dotfill                             & 442.361(1) \\
Pulsar mass function, $f_x$ ($\times 10^{-4} M_{\sun}$)\dotfill  & $\simeq4.8$ \\
Minimum companion mass, $M_c$ ($M_{\sun}$)\dotfill               & $\gtrsim0.1$ \\
\hline
$^1$The mean longitude at 0$^o$ in a circular orbit corresponds \\
 to the ascending node   of the orbit.\\
\end{tabular}

\label{table:data}
\end{table}

\subsection{Colors, light curves and states}

From the Standard~2 data \citep{Jahoda06}, we calculated colors and
intensities with a time resolution of 16 seconds and normalized by
Crab \citep[e.g ][]{Altamirano07}. The PCA observations sample three
different outbursts (see Fig.~\ref{fig:lc}).  The color-color diagrams
show a pattern (not plotted) typical for atoll sources. The power
spectral fits confirm the identification of these states \citep[see
also][]{Kaaret03}.  We looked for kHz QPOs, but found none.

\begin{figure}[!hbtp] 
\resizebox{1\columnwidth}{!}{\rotatebox{0}{\includegraphics{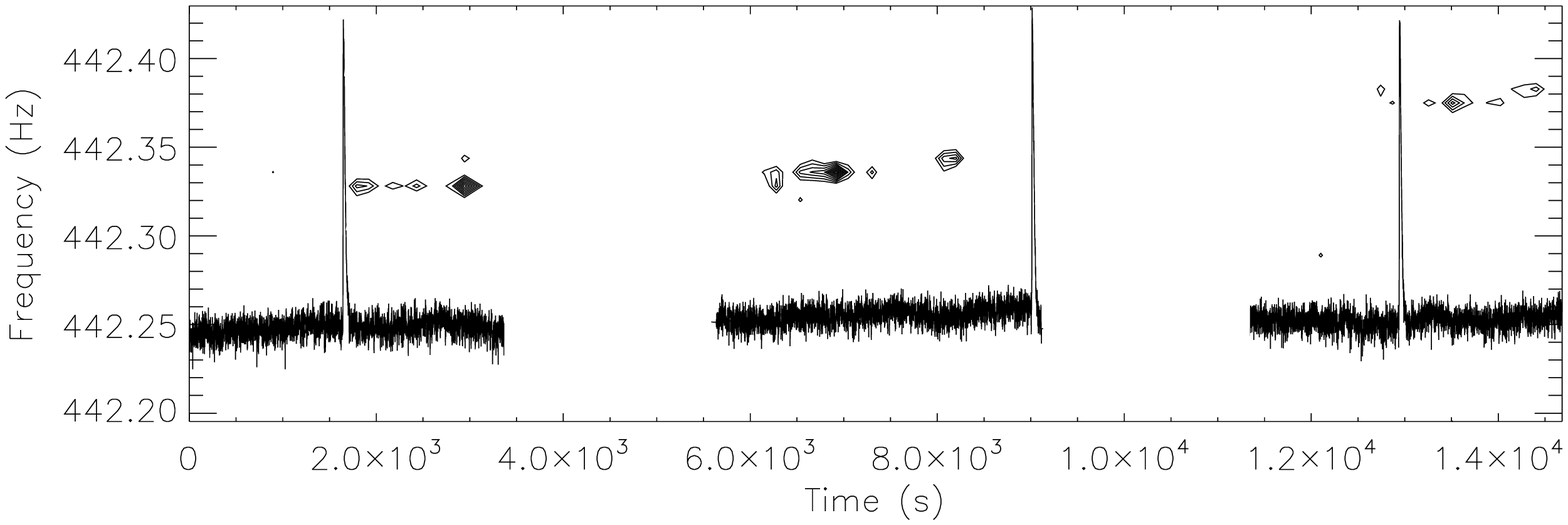}}}
\resizebox{1\columnwidth}{!}{\rotatebox{0}{\includegraphics[clip]{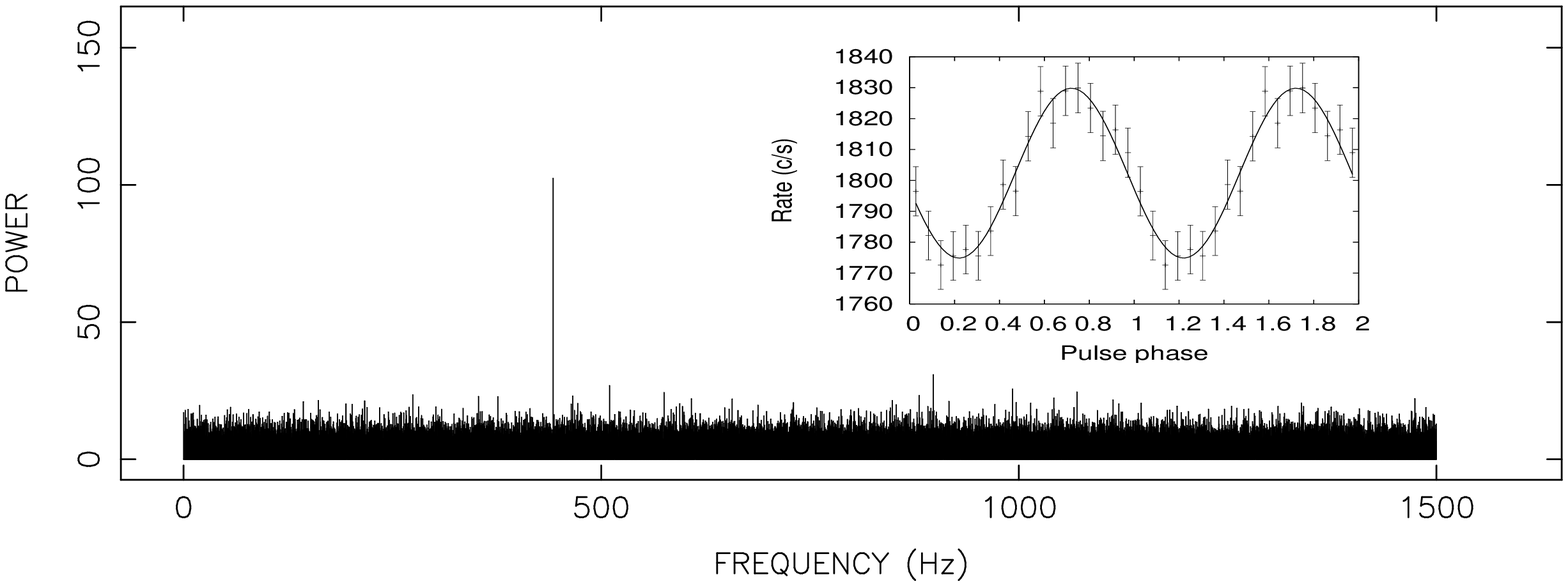}}}
\caption{ \textit{Top:} Dynamical power spectrum of observation
60035-02-03-00 showing intermittent pulsations (contours).  In the
light curve (line) three X-ray bursts are seen. The pulse frequency
drifts due to orbital Doppler modulation. The lowest contour plotted
corresponds to Leahy power $13$ and the highest to 55. The contours
were generated from power spectra for \textit{non}-overlapping 128~sec
intervals of data. \textit{Bottom:} Leahy normalized \citep{Leahy83}
power spectrum of 512~sec of data centered $\sim7$~ksec after the
start of this observation.
Maximum Leahy power is 102, corresponding to a single-trial
probability of $\sim7\cdot10^{-23}$ given Poissonian statistics in the
photon arrival times \citep{vanderklis95}. Inset: The 2--60 keV light
curve folded at the 2.26-ms period. Two cycles are plotted for
clarity. The pulse profile is sinusoidal, with a 95\% upper limit of
0.4\% (rms) on the amplitude of the second harmonic.}
\label{fig:pds}
\end{figure}

No thermonuclear bursts were detected in the first outburst, sixteen
during the second \citep{Kaaret03,Galloway07b} and four during the
third one.
We searched for burst oscillations during all bursts in the
15--4000~Hz frequency range but found none. \citet{Kaaret03} reported
a $\sim4.4\sigma$ burst oscillation at $\sim409.7$~Hz. We find these
authors underestimated the number of trials by a factor of at least
180, as their estimate did not take into account the number of X-ray
bursts analyzed and the fact that a sliding window was used to find
the maximum power. Moreover, we also found that the distribution of
powers is not exponential as these authors assumed. Taking into
account these effects we estimate the significance for their detection
to be $\lesssim2.5\sigma$.

\begin{figure}[!hbtp] 
\resizebox{1\columnwidth}{!}{\rotatebox{-90}{\includegraphics[clip]{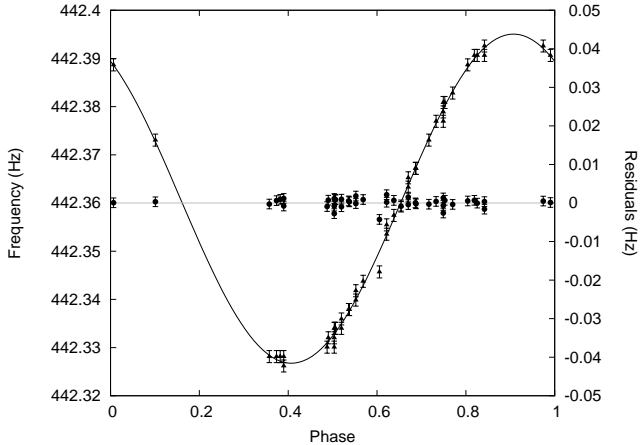}}}
\caption{Pulse frequency as a function of orbital phase.  The
plot has been obtained by folding all data between the first and last
pulse detection in 2001. Pulsations were detected during 6 of the 18
orbital cycles covered.  Drawn curve is the best-fit orbital model,
measured frequencies and post-fit residuals are shown. The residuals'
r.m.s. is $1.2\times10^{-3}$~Hz.} 
\label{fig:phase}
\end{figure}
\subsection{Pulsations}

We inspected each power spectrum for significant features. We
found several, at frequencies $\sim442.3$~Hz in 7 observations:
60035-02-02-04/05/06, 60035-02-03-00/02/03 during the second outburst
and 91050-03-07-00 during the third outburst \citep[see also][for a
detailed analysis of this observation]{Gavriil07}.
\citet{Zand01} concluded that the 1998 and 2001 outbursts from the
LMXB in NGC~6440 were from the same source (Section 2). Since
pulsations are detected in both the 2001 and 2005 outbursts, we can
now conclude that these two outbursts are also from the same
source. Hence, all outbursts observed from NGC~6440 over the last
decade are from SAX J1748.9--2021.
%

The pulsations are detected intermittently, appearing and disappearing
on time scales of hundreds of seconds. The appearance of pulsations
seems to be related to the occurrence of type-I X-ray bursts, but the
relation is not strict. The first two bursts were observed in an
observation on October $8^{th}$ 2001; the first pulsations a day
later. 
During the third outburst we detect four bursts; pulsations were only
detected after the third one. We also detected pulsations with no
preceding burst. The structure of our data does not allow us to tell
if pulsations and/or other bursts occurred during data gaps.
Figure~\ref{fig:pds} (top) illustrates the relation between pulsations
and bursts. The amplitude of the pulsations varies strongly between
$\sim2$\% and (often) undetectable (0.3\% rms amplitude upper limit at
the 95\% confidence level).
Pulsations are seen right after the occurrence of the first and the
third burst, but in the middle of the data pulsations are
present without the detection of a preceding burst (although a burst
could have happened just before the start of this data segment).
In Figure~\ref{fig:pds} (bottom) we show a power spectrum and
corresponding 2--60 keV pulse profile (inset).  In these data the
pulsation is relatively hard; the rms amplitude increases with energy
from $\sim1$\% at 3~keV to $\sim3$\% at 13~keV.

The 2--10~keV luminosity during the observations in which we detected
pulsations was between 3 and $4\times10^{37}$~ergs~s$^{-1}$
\citep[assuming a fixed $N_H=8.2 \times 10^{21}$
cm$^{-2}$;][]{Zand01}.
Other observations at similar flux and those at higher ( up to
$\simeq5\times10^{37}$~ergs~s$^{-1}$ in observation 91050-03-06-00)
and lower fluxes do not show pulsations (see Fig.~\ref{fig:lc}). From
16, 32, 64 and 128~sec average colors we found no significant changes
in the energy spectra correlated with the pulse-amplitude variations.

We studied the pulse frequency drifts using power spectra of 128, 256
and 512~sec data and find a clear 8.7 hours sinusoidal modulation
which we interpret as due to Doppler shifts by binary orbital motion
with that period. In order to obtain an orbital solution, we performed
a $\chi^2$ scan on the orbital parameters using the method described
by \citet{Kirsch04} and \citet{Papitto05}. Our best estimates are
listed in Table~\ref{table:data}. The combination of data gaps and
intermittency of the pulsations yielded aliases, which are taken into
account by the reported errors.
In Figure~\ref{fig:phase} we plot the pulse frequency as a
function of orbital phase.

\section{Discussion}\label{sec:discussion}

We have discovered intermittent pulsations from the neutron-star LMXB
SAX~J1748.9--2021. Pulsations appear and disappear on time scales of
hundreds of seconds. Although we find a suggestive relation between
the appearance of the pulsations and the occurrence of type-I X-ray
bursts (the pulsations appearing after a burst), the relation is
not strict. 
We find bursts with no subsequent pulsations and pulsations with no
preceding burst (although a burst could have occurred in the preceding
data gaps). From the Doppler shifts on the pulsations we determine
that the system is in a near-circular orbit with period of 8.7 hours
and projected radius of 0.39 lightsec.

The stability of the pulsations (after correcting for the binary
orbit) strongly suggests that the pulsation frequency reflects the
neutron star spin frequency and that SAX~J1748.9--2021 is an accreting
millisecond X-ray pulsar.  
The characteristics of the pulsations are reminiscent of the those
found in HETE~J1900.1--2455: in both sources the pulsations were only
intermittently detected and a possible relation between burst
occurrence and pulse amplitude exists.
However, there are differences: in HETE~J1900.1--2455 the pulsations
were only seen during the first two months of the outburst and their
amplitude decreased steadily on timescales of days after the bursts
which might have caused them to reappear \citep{Galloway07a}. In
SAX~J1748.9--2021 we find the pulsations in the middle of the 2001 and
2005 outbursts and not in the beginning. Furthermore, the amplitude of
the pulsations behaves erratically, switching between detection and
non-detection on time scales of hundreds of seconds.  Despite these
differences, the behavior of the pulsations in both sources is so
similar that we consider it likely that the same mechanism causes the
intermittency of the pulsations in both.

\begin{figure}[t] 
\center
\resizebox{1\columnwidth}{!}{\rotatebox{-90}{\includegraphics[clip]{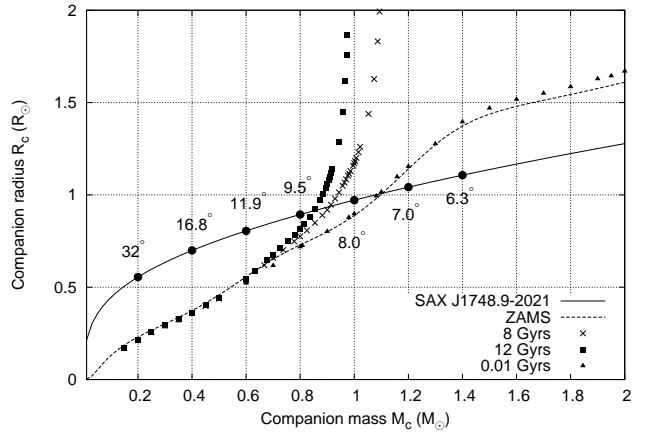}}}
\caption{Mass--radius relationship for a Roche lobe-filling companion
(continuous line), isochrones of 0.01, 8 and 12 Gyrs with solar
metallicity \citep[triangles, crosses and squares, respectively
;][]{Girardi00} and theoretical Zero-age main sequence \citep[ZAMS,
dashed line;][]{Tout96}.  Black circles mark the inclination of the
system as estimated by the mass function of this system.}
\label{fig:rvsm}
\end{figure}

A related system might be Aql~X-1 in which a short-lived ($\sim$150~s)
and very rare (duty cycle of 0.03\%) episode of strong pulsations at
the neutron-star spin frequency has been detected
\citep{Casella07}. In this source, no X-rays bursts were seen in the
$\sim1400$~s before the pulsations, making it unlikely that they were
triggered by a burst. It is unclear if the pulsations in Aql X-1 were
accretion--driven or due to unusual nuclear burning episodes; the same
applies to SAX~J1748.9--2021.
The extreme rarity of the pulsations in Aql X-1 could indicate that
the mechanism behind them is different from that responsible for the
pulsations in HETE~J1900.1--2455 and SAX~J1748.9--2021. Nevertheless,
irrespective of the mechanisms behind the pulsations in these three
sources, it is clear that a strict division between pulsating and
non-pulsating sources cannot be made anymore. It is possible that all
sources pulsate occasionally although the recurrence times could be
very long.

Assuming a constant dipolar magnetic field, following
\citet{Psaltis99} (i.e., assuming a geometrically thin disk and
neglecting inner disk wind mass loss, radiation drag and GR effects)
we estimate the magnetic field to be
$B\gtrsim1.3\times10^{8}$~Gauss. This assumes a 10~km radius
$1.4M_{\odot}$ neutron star and $\dot{M}_{max}$, the highest accretion
rate at which pulses are detected, of 0.28 of the Eddington critical
value as derived from the luminosity observed at the time using a
bolometric flux correction of 1.4 \citep{Migliari06}.  In the standard
magnetic channeling scenario, the question remains of what causes the
large variations in pulse amplitude.

Comparisons between HETE~J1900.1--2455, SAX~J1748.9--2021 and the
other 7 AMSPs can provide clues to understand the pulse-strength
variations.
In SAX~J1748.9--2021 and Aql X-1 the time scales on which the pulse
amplitude can fluctuate are as short as $\sim10^2$~s, too short for
the properties of the neutron star core to change \citep{Galloway07a}.
So, these changes must originate in the disk or the outer layers
of the neutron star envelope. 
\citet{Galloway07a} suggested (for HETE~J1900.1--2455) that the
accumulation of matter on the surface burying the magnetic field
\citep{Cumming01} plays a role.
Our results show that this mechanism probably cannot work for
SAX~J1748.9--2021, as the pulsations are not seen in the beginning of
the outbursts, but instead $\sim3$~weeks and $\sim5$~weeks after the
start of the 2001 and 2005 outbursts, respectively, so after a
considerable amount of matter has already accreted.
Interestingly, we observe pulsations only around a mass accretion rate
of $\simeq2\times10^{-16}\ M_{\odot}/$sec as inferred from the X-ray
luminosity, not above or below, indicating that instantaneous mass
accretion rate rather than total accreted mass is the important
quantity.

In both HETE~J1900.1--2455 and SAX~J1748.9--2021 the pulsations seem
to appear together with bursts although the exact connection is
complex.
This suggests that surface processes may affect the magnetic field.
Hydrodynamic flows in the surface layer of the neutron star may screen
the magnetic field \citep[see review by][and references
within]{Bhattacharya02}; perhaps violent processes like bursts
temporarily affect such flows, diminishing screening and enhancing the
channeling.
%
%
%
%

Alternatively, variations in a scattering or screening medium may 
 cause  the pulse amplitude modulation (see e.g. discussions in
 \citet{Psaltis99}, \citet{Titarchuk02}, \citealt{Gogus07},
 \citealt{Titarchuk07}, \citealt{Casella07} and references within).
For our results, the properties of such a medium should change on
 timescales of hundreds of seconds; note that we did not detect
 spectral changes associated with pulse strength modulation.

With an orbital period of $\sim8.7$ hours, this binary system is
clearly not an ultra-compact binary as usually found in globular
clusters and in fact, SAX~J1748.9--2021 is the AMSP with the longest
orbital period after Aql~X-1, which has an orbital period of
$\sim19$~hrs \citep{Chevalier91,Welsh00}.
The mass-radius relation for a low-mass Roche lobe-filling companion
in a binary \citep{Eggleton83} is $R_c =
0.24~M_{NS}^{1/3}~q^{2/3}~(1+q)^{1/3}~P^{2/3}_{hr} / ( 0.6~q^{2/3} +
\ell og(1+q^{1/3})) $, with $P_{hr}$  the orbital period in hours,
$M_{NS}$ the mass of the neutron star, $R_c$ radius of the companion
and $q=M_c/M_{NS}$, the mass ratio.
Given the mass function and the orbital period and 
assuming a $1.4M_{\odot}$ neutron star, we plot in Figure~\ref{fig:rvsm} the
mass-radius relationship for the companion star.
Given that the age of the Globular Cluster NGC 6440 is
$10\pm2$ Gyrs \citep{Santos04} and its metallicity is approximately
solar \citep{Ortolani94}, in Figure~\ref{fig:rvsm} we also plot the
isochrones for stars with ages of 8 and 12 Gyrs and solar
metallicity. Stars with a $M_c<0.85\rm\,M_{\odot}$ cannot fill the
Roche lobe while
stars with $M_c>0.95\rm\,M_{\odot}$ would have a radius exceeding the
Roche lobe.
This would imply a donor star mass of $0.90\pm0.05M_{\odot}$.
%
However, for masses of 0.95--1.1$\rm\,M_{\odot}$, stars have evolved
off the main sequence so binary mass transfer can have affected the
radius of the donor star, which means we cannot firmly exclude masses
of 0.95--1.1$\rm\,M_{\odot}$.  Therefore a more conservative mass
range for the donor star is 0.85-1.1$\rm\,M_{\odot}$.  Intriguingly,
this requires the inclination to be about $9^o$, which has a
$\lesssim1\%$ a priori probability for an isotropic sample of binary
inclinations.
Of course, this estimate is assuming that SAX~J1748.9--2021 is in a
primordial binary. If a different evolutionary path took place
(e.g. dynamical interactions), the mass of the companion might be much
smaller \citep[see e.g.][]{Zyl04}.


\end{document}